\documentclass{article}

\usepackage{latexsym}
\usepackage{amssymb,amsmath,amsopn}
\usepackage[dvips]{graphicx}
\usepackage{color,epsfig}
\usepackage{url}
\usepackage{multicol}
\usepackage{amsthm}

\usepackage[utf8]{inputenc}

\pdfpagewidth \paperwidth
\pdfpageheight \paperheight
\newcommand*{\twoheadrightarrowtail}{\mathrel{\rightarrowtail\kern-1.9ex\twoheadrightarrow}}

\begin{document}
\title{\textbf{The GraftalLace \\ Cellular Automaton}}
\author{Kaszanyitzky, Andr\'as \\ kaszi75@gmail.com}
\date{}

\maketitle

\textbf{Abstract.} 
We introduce our \emph{GraftalLace Cellular Automaton} in short GLCA which is a new one-dimensional cellular automaton on the regular square lattice. It makes a \emph{monochromatic infinite directed graph} which evolve deterministically row by row, by a defined rule and a single initial row of arc patterns. Arcs overlap each other and partly influence the states of the next cells in another arrangement.
The data structure of GLCA is a \emph{number triangle} or number trapezoid consists of octal digits formed by bit operations. We show examples of GLCA patterns which represent all four classes of Wolfram's classification. Some of the patterns belongs to \emph{Sierpi\'{n}ski-like fractals} as \emph{Pascal Triangle modulo 2} and \emph{modulo 3} patterns which can be realized by GLCA in many different ways. We show these fractals, observe the \emph{reversibility} of the rules and give ideas to extend our automaton by using more colours and other representations to find new interesting patterns.

\maketitle

\section {Definition of GLCA}

I invented GLCA in 1991 inspired by the articles of \emph{Scientific American magazine} about elementary cellular automata of \emph{Stephen Wolfram} [W84,W02] and graftal trees  otherwise recursive fractal plants of \emph{Aristid Lindenmayer} [D86,PL90]. Graftal is a combination of two words: graph + fractal.

GLCA connects root patterns with branch patterns through a junction (otherwise a grid point of the square lattice) by a defined rule. Both patterns are triplets of arcs formed by the incoming and the outgoing arcs of a junction from and into the same horizontal position with the left and right neighbour cells. Usually we get a porous, lace-like chaotic pattern. For illustration see \emph{Figure 1}.

\subsection {Formal definition of GLCA}
GLCA operating with a chain of deterministic finite automata (DFA) and can be represented as a 4-tuple $\langle\Sigma,\phi,B,c_0\rangle$, where $\Sigma$ is an octal alphabet (cell states), $\phi$ is the local transition function, $B$ is a function to define the cell neighbourhood with bit operations and $c_0$ is the initial configuration. GLCA evolves on an array of cells $(s_{l})$ where $l \in \mathbb{N}$ and each cell takes a state from the octal alphabet. This array represents a global configuration $c$, such that $c \in \Sigma^*$. The set of finite configurations of length $l$ is represented as $\Sigma^l$. Cell states in a configuration $c(j)$, where $j \in \mathbb{Z}$, are updated by the next configuration $c(j+1)$ simultaneously by the local transition function $(\phi)$ otherwise the rule $(R)$. This rule tells how to transform all possible octal digits into another octal digit. 

In the next subsections we show two bit operating functions: $B_p(n)$ and $B^q(m)$ to define the neighbourhood of the cells because transition function in GLCA unlike other cellular automata only partly influences the states of the next cells (3 cells, one at the same horizontal position with the left and right neighbour cells) and new states come from another arrangement of the new bit triplets.

The array increases maximum 2 cells in each time steps $(j)$. Evolution of GLCA is represented by a sequence of finite configurations $(c_l)$ given by the global mapping, $\Phi: \Sigma^l \rightarrow \Sigma^{l+2}$.

\begin{figure}[ht]
\centering
\includegraphics[width=0.9\textwidth]{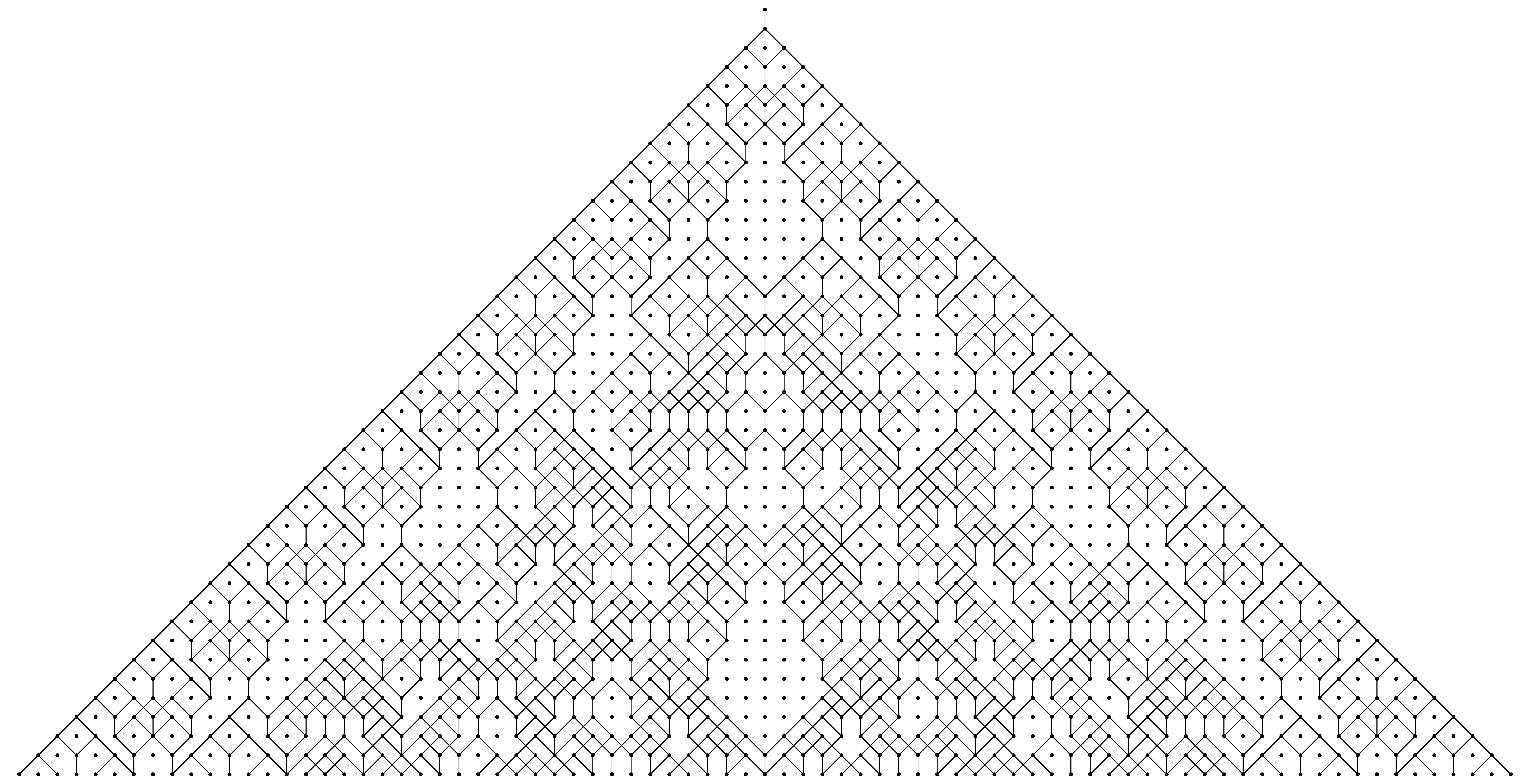}
\centerline{\textbf{Figure 1.}\ First 40 rows of Rule $51254550_{8}$ with gridpoints, from a single}
\centerline{ vertical root. It was the first interesting GLCA pattern I have found in 1991.}
\end{figure}

\[
\begin{picture}(300,40) 

\put(0,20){\line(1,1){20}} 
\put(10,30){\line(0,1){10}} 
\put(20,20){\line(-1,1){20}}

\put(40,20){\line(1,1){10}} 
\put(50,30){\line(0,1){10}} 
\put(50,30){\line(-1,1){10}}

\put(90,30){\line(1,1){10}} 
\put(90,20){\line(0,1){10}} 
\put(90,30){\line(-1,1){10}}

\put(120,20){\line(1,1){10}} 
\put(130,30){\line(0,1){0}} 
\put(140,20){\line(-1,1){20}}

\put(170,30){\line(1,1){10}} 
\put(170,30){\line(0,1){10}} 
\put(180,20){\line(-1,1){10}}

\put(200,20){\line(1,1){10}} 
\put(210,30){\line(0,1){10}} 
\put(220,20){\line(-1,1){10}}

\put(240,20){\line(1,1){20}} 
\put(250,20){\line(0,1){0}} 
\put(260,20){\line(-1,1){10}}

\put(280,20){\line(1,1){0}} 
\put(290,20){\line(0,1){0}} 
\put(300,20){\line(-1,1){0}}

\put(0,20){\circle*{2}}
\put(10,20){\circle*{2}}
\put(20,20){\circle*{2}}
\put(40,20){\circle*{2}}
\put(50,20){\circle*{2}}
\put(60,20){\circle*{2}}
\put(80,20){\circle*{2}}
\put(90,20){\circle*{2}}
\put(100,20){\circle*{2}}
\put(120,20){\circle*{2}}
\put(130,20){\circle*{2}}
\put(140,20){\circle*{2}}
\put(160,20){\circle*{2}}
\put(170,20){\circle*{2}}
\put(180,20){\circle*{2}}
\put(200,20){\circle*{2}}
\put(210,20){\circle*{2}}
\put(220,20){\circle*{2}}
\put(240,20){\circle*{2}}
\put(250,20){\circle*{2}}
\put(260,20){\circle*{2}}
\put(280,20){\circle*{2}}
\put(290,20){\circle*{2}}
\put(300,20){\circle*{2}}

\put(10,30){\circle*{2}}
\put(50,30){\circle*{2}}
\put(90,30){\circle*{2}}
\put(130,30){\circle*{2}}
\put(170,30){\circle*{2}}
\put(210,30){\circle*{2}}
\put(250,30){\circle*{2}}
\put(290,30){\circle*{2}}

\put(0,40){\circle*{2}}
\put(10,40){\circle*{2}}
\put(20,40){\circle*{2}}
\put(40,40){\circle*{2}}
\put(50,40){\circle*{2}}
\put(60,40){\circle*{2}}
\put(80,40){\circle*{2}}
\put(90,40){\circle*{2}}
\put(100,40){\circle*{2}}
\put(120,40){\circle*{2}}
\put(130,40){\circle*{2}}
\put(140,40){\circle*{2}}
\put(160,40){\circle*{2}}
\put(170,40){\circle*{2}}
\put(180,40){\circle*{2}}
\put(200,40){\circle*{2}}
\put(210,40){\circle*{2}}
\put(220,40){\circle*{2}}
\put(240,40){\circle*{2}}
\put(250,40){\circle*{2}}
\put(260,40){\circle*{2}}
\put(280,40){\circle*{2}}
\put(290,40){\circle*{2}}
\put(300,40){\circle*{2}}

\put(7,50){7}
\put(47,50){6}
\put(87,50){5}
\put(127,50){4}
\put(167,50){3}
\put(207,50){2}
\put(247,50){1}
\put(287,50){0}

\put(7,5){5}
\put(47,5){1}
\put(87,5){2}
\put(127,5){5}
\put(167,5){4}
\put(207,5){5}
\put(247,5){5}
\put(287,5){0}

\end{picture}
\]
\centerline{\textbf{Figure 2.}\ A rule as an octal number $R=51254550_8$ means how to connect}
\centerline{all the root patterns (upper triplets) with branch patterns (lower triplets}
\centerline{in reverse order binary code) through the junction (centre point).}

\subsection {Basic definitions}
Our \emph{cell space} is the regular simple upright square lattice. All grid points are cells called the \emph{junctions}. We use the Cartesian coordinate system with upside-down y-coordinates.
We only allow connections between a cell and its closest 3 neighbour cells in a positive (nonzero) vertical direction with arcs and we denote the connection between two cells with a binary digit (1=connected, 0=independent). It means only vertical and diagonal arcs are allowed and the maximum number of the connecting arcs in one junction is six (3 indegrees and 3 outdegrees). We call the possible incoming arrangement of arcs into a junction: the \emph{root pattern}. We call the possible outgoing arrangement of arcs from a junction: the \emph{branch pattern}. We draw only the root pattern (states of the cells) in each time step and upload the next row with bits of the branches by the rule. Branches automatically become roots in another arrangement in the next time step.

Both patterns form binary triplets which have 8 possible variants denoted by an octal digit therefore we use the octal \emph{alphabet}: $\Sigma={\{0,1,2,3,4,5,6,7\}}$. The \emph{state} of a cell ($s$) specified by its incoming arc triplet, the root pattern. By choosing an eight-digits long octal number ($R$) we get a \emph{rule} for our GLCA which tells how to combine the potential root patterns with branch patterns. This rule gives the \emph{local transition function} of GLCA: $\phi(s)$. Root patterns are denoted by the place values of the rule, their connecting branch patterns are denoted by the digits of the rule. See \emph{mini trees} on \emph{Figure 2} and more details on \emph{Figure 5}.

For practical reasons we denote the branch patterns with a \emph{reverse order binary number} because in the next time step (next row of the evolving pattern) branch arcs become root arcs in another arrangement where every arc belongs to different junctions in a reverse order. Both patterns overlap each other. \emph{Figure 3} shows the overlapping patterns and the potential adjacency of neighbour cells.

\[
\begin{picture}(135,75) 

\put(12,12){\line(1,1){21}} 
\put(42,42){\line(1,1){21}}
\put(42,12){\line(1,1){21}} 
\put(72,42){\line(1,1){21}}
\put(72,12){\line(1,1){21}} 
\put(102,42){\line(1,1){21}}

\put(37.5,13.5){\line(0,1){18}} 
\put(37.5,43.5){\line(0,1){18}}
\put(67.5,13.5){\line(0,1){18}} 
\put(67.5,43.5){\line(0,1){18}}
\put(97.5,13.5){\line(0,1){18}} 
\put(97.5,43.5){\line(0,1){18}}

\put(12,63){\line(1,-1){21}}
\put(42,33){\line(1,-1){21}}
\put(42,63){\line(1,-1){21}}
\put(72,33){\line(1,-1){21}}
\put(72,63){\line(1,-1){21}}
\put(102,33){\line(1,-1){21}}

\put(7.5,7.5){\circle{12}}
\put(37.5,7.5){\circle{12}}
\put(67.5,7.5){\circle{12}}
\put(97.5,7.5){\circle{12}}
\put(127.5,7.5){\circle{12}}

\put(37.5,37.5){\circle{12}}
\put(67.5,37.5){\circle{12}}
\put(97.5,37.5){\circle{12}}

\put(7.5,67.5){\circle{12}}
\put(37.5,67.5){\circle{12}}
\put(67.5,67.5){\circle{12}}
\put(97.5,67.5){\circle{12}}
\put(127.5,67.5){\circle{12}}

\put(34,34){X}
\put(64,34){Y}
\put(94.5,34){Z}
\put(65,4){S}

\end{picture}
\]
\centerline{\textbf{Figure 3.}\ Overlapping roots and branches. We show 3 cells}
\centerline{in the middle row (X,Y,Z) with all of their possible connections.}
\centerline{Their branches make a new root pattern by the incoming arcs of cell S.}

\bigskip

\[
\begin{picture}(160,70) 

\linethickness{0.6mm}
\put(34.5,65){\line(1,0){91}}   
\put(4.5,35){\line(1,0){151}}
\put(4.5,5){\line(1,0){151}}

\put(5,4.5){\line(0,1){31}}
\put(35,4.5){\line(0,1){61}}
\put(65,4.5){\line(0,1){61}}
\put(95,4.5){\line(0,1){61}}
\put(125,4.5){\line(0,1){61}}
\put(155,4.5){\line(0,1){31}}

\linethickness{0.2mm}
\put(15,5){\line(0,1){30}}
\put(25,5){\line(0,1){30}}
\put(45,5){\line(0,1){30}}
\put(55,5){\line(0,1){30}}
\put(75,5){\line(0,1){30}}
\put(85,5){\line(0,1){30}}
\put(105,5){\line(0,1){30}}
\put(115,5){\line(0,1){30}}
\put(135,5){\line(0,1){30}}
\put(145,5){\line(0,1){30}}

\put(47,46){\makebox{X}}
\put(77,46){\makebox{Y}}
\put(107,46){\makebox{Z}}

\put(26.5,17.5){\makebox{\tiny{$x_0$}}}
\put(47,17.5){\makebox{\tiny{$x_1$}}}
\put(67,17.5){\makebox{\tiny{$x_2$}}}

\put(57,17.5){\makebox{\tiny{$y_0$}}}
\put(77,17.5){\makebox{\tiny{$y_1$}}}
\put(97,17.5){\makebox{\tiny{$y_2$}}}

\put(85.5,17.5){\makebox{\tiny{$z_0$}}}
\put(107,17.5){\makebox{\tiny{$z_1$}}}
\put(127,17.5){\makebox{\tiny{$z_2$}}}

\end{picture}
\]
\centerline{\textbf {Figure 4.}\ Data representation of overlapping branch patterns}
\centerline{from 3 junctions (X,Y,Z above) into 5 next junctions (below).}
\centerline{Binary digits in a new combination represent a new root pattern: $S=\overline{x_2y_1z_0}$.}

\bigskip

The rule defines the corresponding \emph{branch pattern} for any potential root patterns otherwise for any potential states of the cells: $s_{i,j} \rightarrow \phi(s_{i,j})$ which means 3 possible outgoing connecting arcs from the junction. These arcs form an octal digit as a reverse order binary triplet by the highest bit: $(v_{i,j} \rightarrow v_{i+1,j+1})$, the middle bit: $(v_{i,j} \rightarrow v_{i,j+1})$ and the lowest bit: $(v_{i,j} \rightarrow v_{i-1,j+1})$ where $v$ is a cell otherwise a grid point (vertex). The horizontal position of the vertex in a row denoted by $i$, its vertical position otherwise the actual time step is denoted by $j$ where $i,j \in \mathbb{Z}$.

One branch pattern (outgoing triplet of arcs from a junction) partly influences the states of its 3 different neighbour cells in the next row by changing their corresponding bits. The states of the cells come from their \emph{root pattern} as their incoming connecting arcs from 3 different cells into a junction by the highest bit: $(v_{i-1,j} \rightarrow v_{i,j+1})$, the middle bit:  $(v_{i,j} \rightarrow v_{i,j+1})$ and the lowest bit: $(v_{i+1,j} \rightarrow v_{i,j+1})$.

Our rule is assigning all possible $s$ values into not necessarily different $\phi(s)$ values. The total number of the possible rules are $8^{8}$=$16777216$. We avoid growing branches from nothing therefore the last digit of the rule is always equal to zero. The number of the remaining rules is $8^{7}$=$2097152$. 

\subsection{Bit operating functions}

In this section we show how the local transition function creates the states of the new cells in the next time step automatically (1). We define a new bit operating function  $B_p(n)$ which gives back the value of the $2^p$ component of an octal digit $n$.

For example: $B_2(6)=4, \qquad B_1(6)=2, \qquad B_0(6)=0$.

\begin{equation}
\phi: s_{i,j+1}=\sum_{p=0}^{2}B_p(\phi(s_{i+1-p,j}))
\end{equation}

Let's consider another bit operating function $B^q(m)=b$ which means let the $q$th bit of the octal number $m$ is equal to bit $b$. For example: if $m=0$ then $B^2(m)=1$ means $m=4$, and after that $B^0(m)=1$ means $m=5$. Now we can show how the local transition function creates the branches (2) at the same time with root patterns. It is only another grouping of the arcs.

The following branches partly influence the states of 3 different cells in the next row:

\begin{equation}
s_{i,j} \rightarrow \phi(s_{i,j})=
\begin{cases}
B^2(s_{i+1,j+1})=B_2(\phi(s_{i,j}))\\
B^1(s_{i,j+1})=B_1(\phi(s_{i,j}))\\
B^0(s_{i-1,j+1})=B_0(\phi(s_{i,j}))
\end{cases}
\end{equation}

\bigskip

See \emph{Figure 4} for data representation of a new root pattern (new state of the cell S below cell Y) made by combined bits of different branch patterns. Branch patterns with reverse order bits: $X \rightarrow \overline{x_2x_1x_0}$, $Y \rightarrow \overline{y_2y_1y_0}$, $Z \rightarrow \overline{z_2z_1z_0}$ automatically make a new root pattern in the right order: $S=\overline{x_2y_1z_0}$.

\begin{figure}[ht]
\centering
\includegraphics[width=0.9\textwidth]{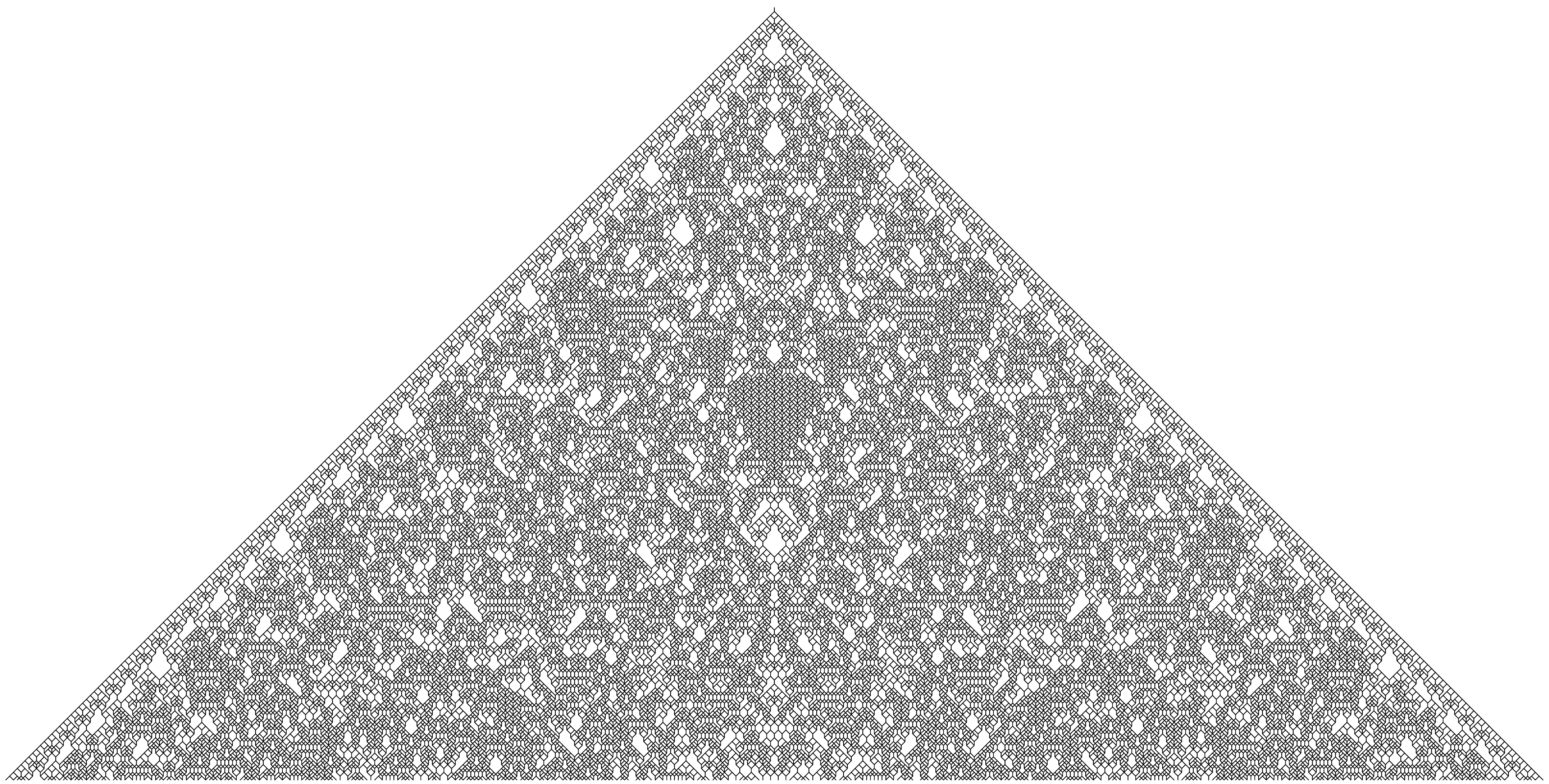}
\centerline{\textbf{Figure 5.}\ Rule $51254550_8$, 200 rows.}
\centerline{}
\includegraphics[width=0.9\textwidth]{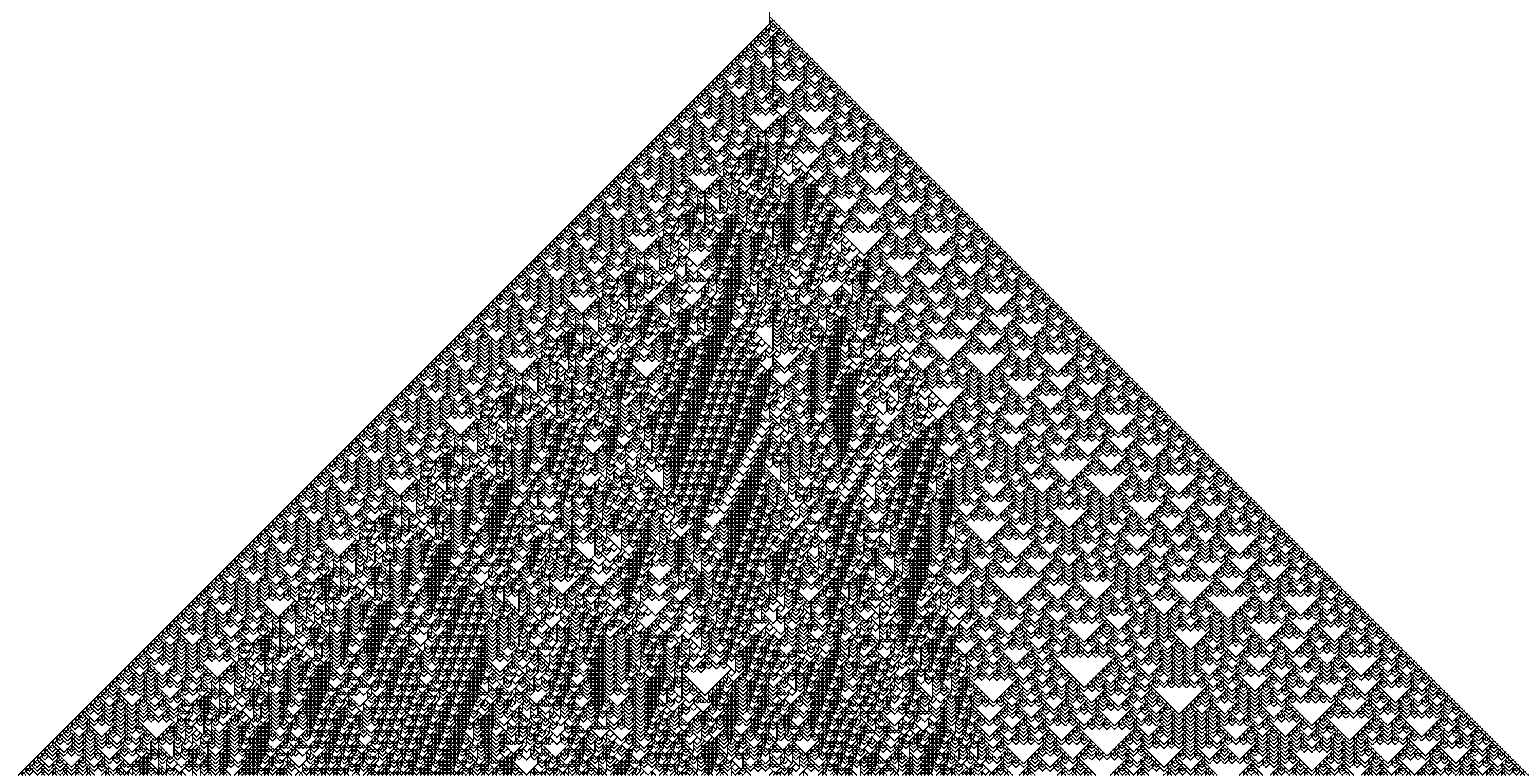}
\centerline{\textbf{Figure 6.}\ Complex pattern of GLCA, Rule $71055670_8$, 200 rows.}
\end{figure}

\section{Symmetric fractal patterns}
We can find all the pattern groups of Wolfram's classification (Class I-IV) among GLCA patterns otherwise the evolution of the patterns leads to homogenous, regular, chaotic and complex patterns. See \emph{Figure 5} and \emph{6}.

We show how can we realize \emph{Pascal triangle modulo $3$} symmetric fractal pattern by the monochromatic GLCA. See \emph{Figure 7}.

\begin{figure}[ht]
\centering
\includegraphics[width=0.54\textwidth]{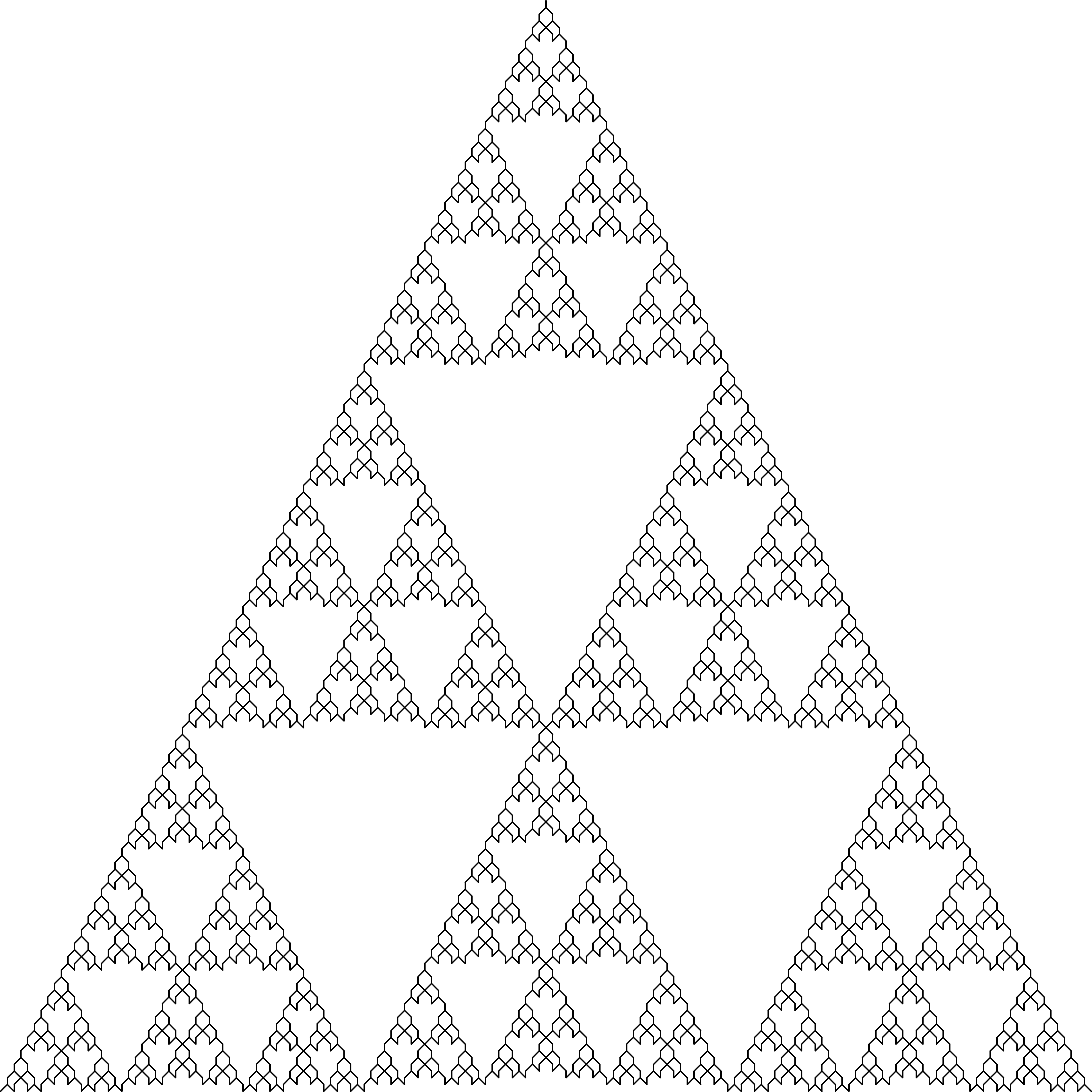}
\centerline{}
\centerline{\textbf{Figure 7.}\ Rule $00520520_8$ of GLCA = Pascal triangle modulo 3.}
\centerline{162 rows of a nested pattern = 4th approximation ($2\cdot3^n$ rows of arcs).}
\end{figure}

The \emph{Sierpi\'{n}ski triangle} (Pascal triangle modulo 2) pattern can be realized in many ways. For example by applying an XOR binary operation or an iterated function system (IFS) rule onto a binary square matrix. We get the same result with Wolfram's elementary cellular automaton [WE]. His simple rules 60, 102, 90 and 126 also give this pattern on different ways. 

GLCA also gives other possibilities to realize this fractal pattern. The simplest one, Rule $00050550_8$ can be drawn from any single root arc. The rule means draw two vertical branch arcs from single arcs and do not draw in other cases. See \emph{Figure 8}. Rule $00020520_8$, $06523520_8$ and $00720520_8$ also make this fractal in another way.

\begin{figure}[ht]
\centering
\includegraphics[width=0.39\textwidth]{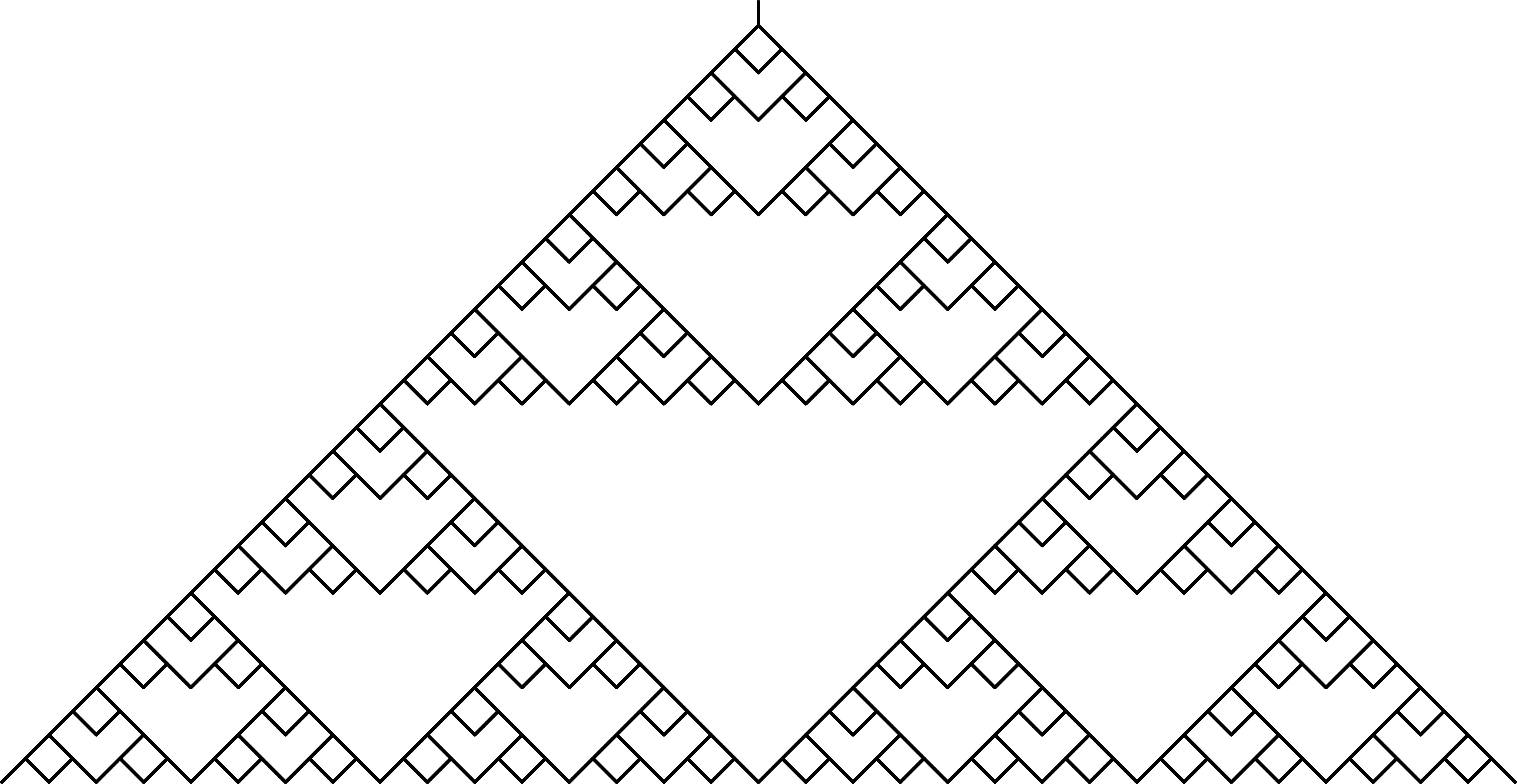}
\centerline{}
\centerline{\textbf{Figure 8.}\ Rule $00050550_8$ of GLCA = Pascal triangle modulo 2.}
\centerline{33 rows of a nested pattern = 5th approximation ($1+2^n$ rows of arcs).}
\end{figure}

We can realize \emph{Pascal triangle modulo 3} pattern in many different ways also. As an IFS fractal [W02], by recursive curves otherwise by Hamiltonian paths or Hamiltonian cycles [K17a,K17b]. With Wolfram's automaton we have to use more colours [W84,W02] (3 colours, totalistic rule, code 420) unlike my monochromatic GLCA pattern on \emph{Figure 7}.

\section {Searching for reversible rules}

A reversible cellular automaton is a system that is deterministic in both directions in terms of time. It is also called invertible cellular automaton. In GLCA it means if we change the direction of all arcs of the mini trees into reverse we can continue the drawing at the other side of a root pattern. Most of the cases these directions belongs to different rule numbers. For reversible rules we have to find bijective pairs with the same rule number.

In reversible rules we have to avoid growing branches from nothing therefore the last digit of the rule is always equal to zero. We have to use assignments amongst root patterns and branch patterns with one-to-one correspondence. We have 7 different patterns so the maximum number of these unambigous assignments are equal to the number of the permutations of our patterns: $7! = 5040$.

By leaving odd numbers of digits at their place-values in the octal rule number and changing the remaining digits pairwise by mutuality of the number of the place value and the correlating digit we get the following sum of binomials: ${7}\choose{0}$+${7}\choose{2}$+$3$${7}\choose{4}$+$15$${7}\choose{6}$. In this case the rule number does not depend on the direction of the mini trees (assignments) therefore we get 232 different reversible rules. This is the number of the self-inverse permutations on 7 letters, also known as involutions [OEIS]. For example Rule $67234510_8$ is a reversible one.

In the remaining cases we get a different rule number by changing the direction of the drawing. We get the correlating rule pair by replacing the digit values with the place-values of the rule number for example: Rule $35724160_8$ and Rule $51637420_8$ are correlating pairs.

\section {Extensions and variations of the basic idea} 
By using the same 3 arcs long root and branch patterns and bichromatic arcs (2 drawing colours and 1 background colour)
the triplets can be described as 3-digit long numbers in ternary numeral system. In this case we combine $3^{3}$ root patterns with also $27$ branch patterns and we have $27^{27}=3^{81}$ different rules. We can represent these numbers with $0$ to $9$ and {A} to {Q} symbols (as digits of numeral system 27). In this case the rule is a $27$-digit long number consisting of these symbols. See \emph{Figure 9}.

We recommend Wolfram's method the \emph{totalistic rules} to define the assignments in an easier way. Instead of defining branch patterns for every possible root pattern it is enough to assign branch patterns to groups of root patterns as hues or densities of arcs. These hues or densities are equal to the sum of the digits of a root pattern. For example in monochromatic GLCA (1 drawing and 1 background colour) we have binary triplets as root patterns. The sum of the digits is between 0 and 3 therefore it is enough to define 4 assignings instead of 8. By using bichromatic arcs (2 drawing and 1 background colour) we have ternary triplets as root patterns. The sum of the digits is between 0 and 6 therefore it is enough to define 7 assignings instead of 27.

\begin{figure}[ht]
\centering
\includegraphics[width=1.0\textwidth]{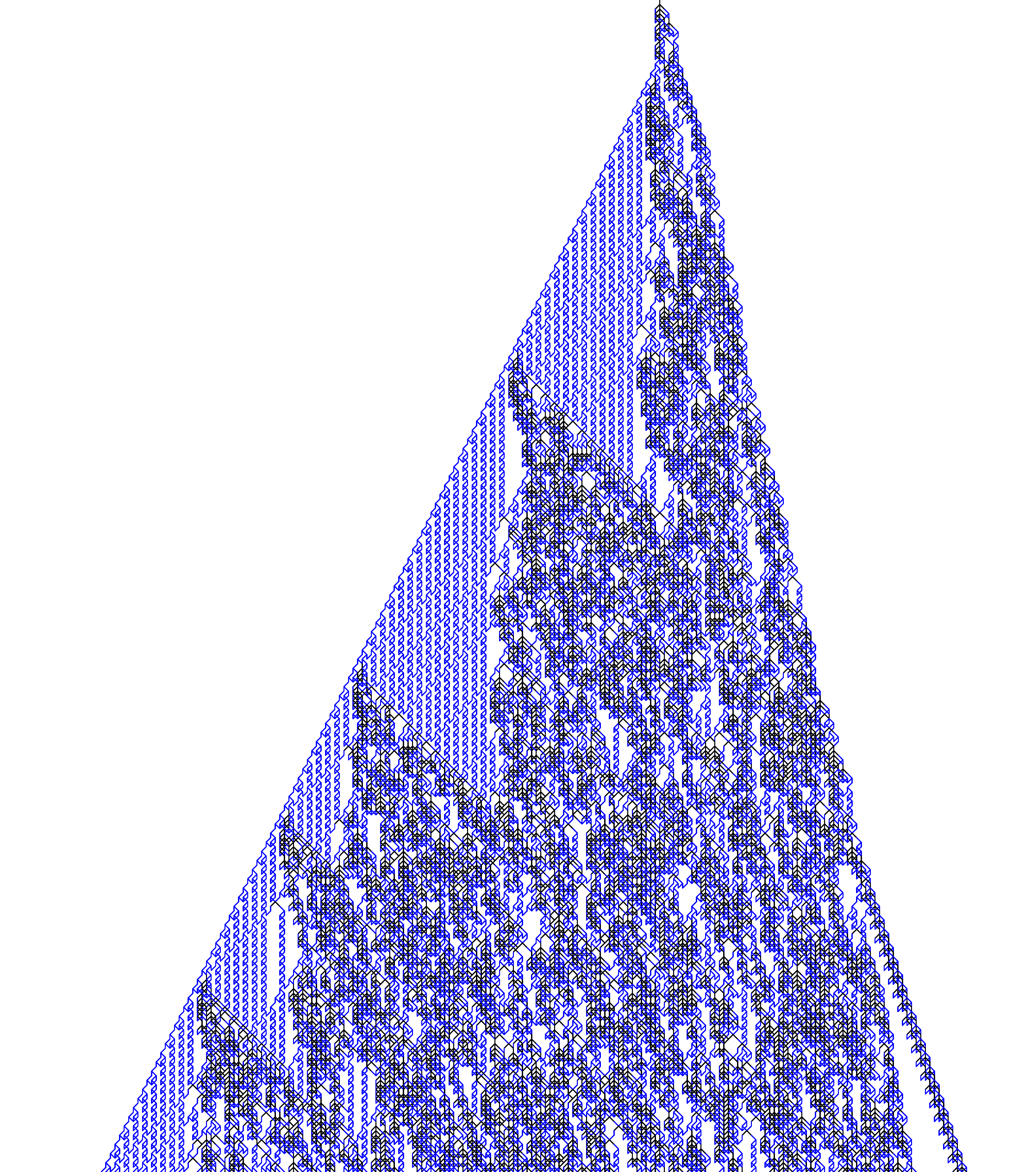}
\centerline{}
\centerline{\textbf{Figure 9.}\ Rule $HPD8962896DGH067K4MHQL013C0_{27}$}
\centerline{Complex pattern in bichromatic version of GLCA.}
\centerline{300 rows, root pattern is a single black vertical arc.}
\end{figure}

\clearpage

We can imagine GLCA on a fixed width space or on a cylindric grid also.

We can colour and draw only the junctions instead of the arcs. It is a number triangle (from a single root pattern) or a number trapezoid (from a single row of root patterns). We have to use 8 colours to show the root patterns (junctions as data containers contain these octal values). Rule $51254550_8$ on \emph{Figure 1} and \emph{Figure 5} makes the following \emph{number triangle} of octal digits: $2, 104, 10504, 1042104,$ $105154504, 10430706104, ...$ etc.

The number triangle constructed as follows: for example root patterns $7,4,2,1$ are assigned with branch pattern $5$. As a reverse binary number it is equal to $1+0+4$. From root pattern $2$ we get $104$, then from root pattern $104$ we get:

\medskip
\centerline{\underline{  1 0 4  }}
\centerline{1 0 4 . .}
\centerline{. 0 0 0 .}
\centerline{\underline{. . 1 0 4}}
\centerline{1 0 5 0 4}
\medskip

We can make the 2D version of GLCA represented by arcs in 3D cubic space. Consider y axis as the growing direction otherwise the time. We get 9 possible arcs (1 vertical, 4 diagonal and 4 space diagonal arcs) at every grid point. By using monochromatic arcs we get $2^{9}$ possible root patterns and the same $512$ different branch patterns. It's worth to use a grouping of arcs to define the rules in an easy way. By using totalistic rules we have to summarize the number of arcs in an elementary pattern. In this case a pattern consists of 0 to 9 arcs therefore a totalistic rule contains 10 numbers between $0$ and $511$. These numbers symbolize a branch pattern for each density or hue of arcs. Beyond the natural monochromatic representation we can visualize every layer as a 2D animation as patterns change in time steps like a stroboscope. In this case it's worth to represent the coloured junctions as a square tessellation instead of the arcs. It could be a closer relative of \emph{Conway's Game of Life}.

\section {Summary}
We have introduced our GraftalLace Cellular Automaton which makes a one-dimensional infinite monochromatic digraph otherwise an octal number triangle or number trapezoid by partly influences the states of the neighbour cells with bit operations. We have shown new ways to make known symmetric fractal patterns and unknown complex patterns. The monochromatic GLCA has $8^8$ possible rules. We have chosen $8^7$ rules in which none of the branches grow from nothing. We have found $7!$ unambigous rules in which $232$ are reversible. We have shown possibilities to represent and extend our automaton in different ways. The 2D version of GLCA can be represented by a 3D graph in cubic space or as a 2D animated tessellation formed by the coloured junctions changing in time steps. It could be a closer relative of \emph{Conway's Game of Life}. Beyond cryptographic utilization, the physical, chemical and biological connections might also be interesting.

\clearpage \noindent \textbf{References.}

\smallskip \noindent [W84] Wolfram, S.: \emph{Computer Software in Science and Mathematics}, Scientific American, Vol. 251, Issue 3, September 1984.
\smallskip

\smallskip \noindent [W02] Wolfram, S.: \emph{A New Kind of Science},
Wolfram Media Inc., 2002
\smallskip

\smallskip \noindent [D86] Dewdney, A. K.: \emph{Computer Recreations --- of fractal mountains, graftal plants and other computer graphics at Pixar}, Scientific American, Dec. 1986
\smallskip

\smallskip \noindent [PL90] Prusinkiewicz, P.\ and Lindenmayer, A.: \emph{The algorithmic beauty of plants}, Springer, 1990.

\smallskip \noindent [WE] \emph{Wolfram MathWorld / Elementary Cellular Automaton}, \\ \url{http://mathworld.wolfram.com/ElementaryCellularAutomaton.html}
\smallskip

\smallskip \noindent [K17a] Kaszanyitzky, A.: \emph{The generalized Sierpi\'{n}ski Arrowhead Curve}, 2017 \\
\url{https://arxiv.org/abs/1710.08480}
\smallskip

\smallskip \noindent [K17b] Kaszanyitzky, A.: \emph{Triangular fractal approximating graphs and their covering paths and cycles}, 2017 \\
\url{https://arxiv.org/abs/1710.09475}
\smallskip

\smallskip \noindent [OEIS] Sloane, N.J.A.: \emph{The On-line Encyclopedia of Integer Sequences}, 
\\
\url{http://oeis.org/A000085}
\smallskip

\end{document}